\documentclass[12pt]{article}
\usepackage[super,compress]{cite}
\usepackage{graphicx}
\usepackage{braket}
\usepackage{epsfig}

\newcommand{\be}{\begin{equation}}
\newcommand{\ee}{\end{equation}}

\newcommand{\bea}{\begin{eqnarray}}
\newcommand{\eea}{\end{eqnarray}}
\newcommand{\beq}{\begin{equation}}
\newcommand{\eeq}{\end{equation}}
\newcommand{\nn}{\nonumber}

\def\fun#1#2{\lower3.6pt\vbox{\baselineskip0pt\lineskip.9pt
\ialign{$\mathsurround=0pt#1\hfil##\hfil$\crcr#2\crcr\sim\crcr}}}

\begin{document}

\title{Propagators of resonances  and rescatterings of the decay products}

\author{A.V. Anisovich$^{+}$, V.V. Anisovich$^+$, M.A.
Matveev$^+$, Nyiri$^*$, \\ A.V. Sarantsev$^{+}$,  A.N. Semenova$^+$,
}
\maketitle

\begin{center}
{\it
$^+$National Research Centre ''Kurchatov Institute'':
Petersburg Nuclear Physics Institute, Gatchina, 188300, Russia}


{\it $^*$Institute for Particle and Nuclear Physics, Wigner RCP,
Budapest 1121, Hungary}

\end{center}

\begin{abstract}

Hadronic resonance propagators which take into account the
analytical properties of decay processes are built
in terms of the dispersion relation technique. Such propagators
can describe multi-component systems, for example, those when quark
degrees of freedom create a resonance state, and decay products correct
the corresponding pole by adding hadronic deuteron-like components. Meson
and baryon states are considered, examples of particles with different
spins are presented.
\end{abstract}

Keywords: Quark model; resonance; exotic states.

PACS numbers: 12.40.Yx, 12.39.-x, 14.40.Lb

\section{Introduction}

Nowadays we face a pressing request for studying multi-component systems,
in particular, those with concurrent parts of quark and hadron degrees of
freedom.
Recent experimental evidences for exotic states
(see Refs.~[\citen{LHCbReview,LHCb,pdgKS}] and references therein) definitely
indicate the important role of both short-distance physics (predominantly
quark-gluon one) and long-distance hadron physics where the notion of
deuteron-like systems or molecules looks quite appropriate \cite{VO,DRGG}.
The active discussion of the pentaquark topic is in line with this trend
\cite{1603.02376,1602.07069,1602.06791,1602.02433,1601.02092,
lebed-D92(2015)114030,stone,Cheng-D92(2015)086009,burns,maiani,as}.

The two-component structure of resonances can reveal itself in propagators
of the resonances. A corresponding consideration of meson resonances is
performed in Ref.~[\citen{AMSS_tetra}] for tetraquark systems with hidden
charm where meson states for decay processes were taken
into account (but with non-relativistic spin wave functions). In
this paper we present the relativistic consideration of both meson and
baryon systems with spins. An important point in this consideration is to
keep the analytic amplitude with correct singular structure.

Let us turn to a standard description of resonances.
The Breit-Wigner pole \cite{bre-wig} gives us a description of a resonance
state in a form of particle propagator, for non-relativistic and
relativistic cases the pole amplitudes read:
\bea  \label{1}
     {\rm non-relativistic}:&&\qquad
     \frac{G_{\imath}G_{\jmath}}{E_0-E-i\frac{\Gamma}{2}}\,, \\
{\rm relativistic}:&&\qquad      \frac{G_\imath G_\jmath}{M^2-s-i\Gamma M
} \,.
\nonumber
    \eea
Here $E$ is the energy of the non-relativistic system, and $E_0$ is the
energy of the resonance level; $\Gamma$ is the width of the resonance
and $G_{\imath},\,G_{\jmath}$ are couplings with initial and final states. For
the relativistic case the total energy $\sqrt s$ includes the mass of the
system and $M$ refers to the resonance mass.

\begin{figure}
\centerline{\epsfig{file=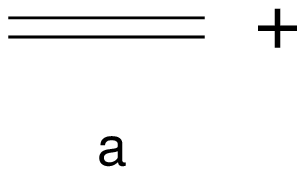,width=0.30\textwidth}\hspace{-5mm}
                \epsfig{file=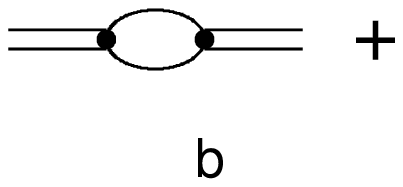,width=0.30\textwidth}
                \epsfig{file=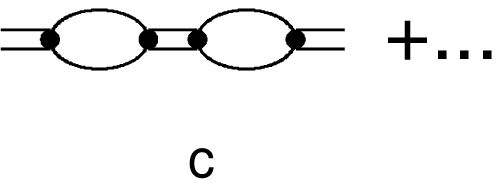,width=0.30\textwidth}}
\vspace{-10mm}
\caption{Graphic representation of the Breit-Wigner
pole term as an infinite set of transitions $resonance\,state\to  decay\;
products\to resonance\,state$.
\label{f1}}
\end{figure}

The energy independent width corresponds to a rough approximation, for
the study of the $\pi p$ scattering in the $\Delta(1240)$ region (hadrons
$\pi p$ are in the $P$-wave) Gell-Mann and Watson \cite{gel-wat} suggested
to use the energy dependent width:
\be \label{2}
\Gamma \to \gamma\frac{k_{\pi p}^3}{1+R^2k_{\pi p}^2} \,,
\ee
where $k_{\pi p}$ is the relative momentum of $\pi p$ in the c.m. system.
Actually the width in the form Eq. (\ref{2}) takes into account
the threshold singularity ($k^{2L+1}$ where $L=1$ is the orbital momentum)
inherent to transitions $\Delta\to\pi p\to \Delta $.
The threshold singularity appears when we consider a set of diagrams
related to decay processes such as shown in Fig. {\ref{f1}}. But Eq. (\ref{2})
contains also singularities which are absent in the scattering amplitude.

First, there are those related to
\be  \label{3}
k_{\pi p}=\sqrt{\frac{[s-(m_p+m_\pi)^2][s-(m_p-m_\pi)^2]}{4s}},
\ee
namely, the square root singularities at $s=0$ and $s=(m_p-m_\pi)^2$;
the form factor $1/[1+R^2k_{\pi p}^2]$ has a pole singularity which gives
zeros in the related amplitudes.
For the precise use of amplitudes with the resonance propagators one
needs to take into account the contributions of decay processes without
false singularities, i.e. to use the corresponding loop diagrams.

The paper is assembled as follows.
In Section 2 spinless hadron resonances are given and multi-channel cases
are considered. Sections 3 and 4 are devoted to particles with spins,
correspondingly, to meson and baryon resonances. In Section 5 we
investigate the problem of formation of the deuteron-like components in
states belonging originally to quark-gluon ones. In Appendices A, B, C
elements of technique for working with loop diagrams of spin particles are
given.

\section{Propagators for meson resonances}

For the inclusion of the loop diagrams into the resonance propagator the
D-function technique is appropriate, we use it here.
First, we consider scalar mesons, after that we generalize the
consideration to cases of spin particles.

\subsection{Loop diagrams in the resonance propagator}

The set of diagrams of Fig.~\ref{f1} reads:
   \bea
\label{4}
&&
\frac{1}{-s+m^2}+\frac{1}{-s+m^2}B(s)\frac{1}{-s+m^2}+
\frac{1}{-s+m^2}B(s)\frac{1}{-s+m^2}B(s)\frac{1}{-s+m^2}+...
\nonumber\\
&& \qquad =\frac{1}{-s+m^2-B(s)}\, ,
\eea
     where $B(s)$ is the contribution of the loop diagram related to the
resonance decay (in Fig. \ref{f1} it is supposed that we deal with a
two-particle decay). If a resonance state decays into several channels,
one should replace:
\be \label{5}
     B(s)\to \sum \limits_{\ell=1}^n B^{(\ell)}(s)
\ee
     where $n$ is the number of open channels. In the standard
Breit$-$Wigner approach the $s$-independent loop diagrams are used:
\be \label{6}
      M^2 =m^2-\sum \limits_{\ell=1}^n Re B^{(\ell)}(M^2),\quad M\Gamma=
\sum \limits_{\ell=1}^n Im B^{(\ell)}(M^2)\,.
\ee
Following the Gell-Mann$-$Watson prescription \cite{gel-wat} one
takes into account the $s$-dependent imaginary part of the loop diagrams:
\be
\label{7}
Im B^{(\ell)}(s)= \rho_\ell (s) g_\ell^2(s),\qquad
\rho_\ell (s)=\frac{k_\ell^{2L_\ell+1}}{8\pi\sqrt{s}},
\ee
     where $\rho_\ell (s)$ is the phase space for loop-diagram particles,
$ g_\ell(s)$ is the vertex for the transition
{\it resonance state$ \to $decay particles of the $\ell$-state}, and
$L_\ell$ is the orbital momentum of particles in the loop diagram.
But, as it was discussed above, the imaginary part alone contains false
singularities.
For reproducing the analytical amplitude correctly, one needs to take into
account the real part of $B^{(\ell)}(s)$ as well.

\subsection{ D-function and loop diagrams with $L=0$ for the one-pole
amplitude}

Let us consider a loop diagram above the threshold, at $s>(M_a+M_b)^2$.
The equation for one-pole and one-channel D-function reads:
\bea \label{8}
D=d+D\,B_{}\,d_ ,
\quad d=\frac{1}{m^2-s},\quad
B_{}=\int\limits_{(M_a+M_b)^2}^\infty\frac{ds'}{\pi}
\frac{G^2\rho(s') }{s'-s-i0}.
\eea
Here $m$ is a bare mass of this state, the factor $B_{}$ goes from the
loop diagram formed by hadrons ($a,b$), and $M_a,M_b$ are masses of the
loop mesons. The phase space factor for the $S$-wave state ($L=0$) is:

\be \label{9}
\rho_{}(s)=\frac{\sqrt{[s\!-\!(M_a\!+\!M_b)^2][s\!-\!(M_a\!-\!M_b)^2]}}{16\pi
s}\,.
\ee
The convergency of the integral for $B(s)$ can be organized either due to
introducing a $s$-dependence of the vertex $G\to G(s')$ or by switching
the subtraction procedure:
\bea \label{10}
&&B(s)=\int\limits_{(M_a+M_b)^2}^\infty\frac{ds'}{\pi} \cdot\frac{G^2\,\rho(s')}{s'-s-i0}\,\,\to
\nn \\
&&
b_0+\int\limits_{(M_a+M_b)^2}^\infty
\!\!\!\!\!\!\frac{ds'}{\pi}
\frac{\big[s\!-\!(M_a\!+\!M_b)^2\big]\cdot G^2\,\rho(s')}{(s'\!-\!(M_a\!+ \!M_b)^2) (s'\!-\!s\!-i0)}
    \,.
\eea
Imposing $G^2=1$ we write for positive $s$, $s>(M_a+M_b)^2$:
\bea \label{11}
B_{}(s)&=&b_0+ \beta\frac{(M_a+M_b)^2-s}{s(M_a+M_b)^2}
+\frac{\sqrt{[s-(M_a+M_b)^2][s-(M_a-M_b)^2]}}{16\pi s}\times
\nn \\ &&
\bigg[\frac{1}{\pi}\ln
\frac{\sqrt{s-(M_a-M_b)^2}-\sqrt{s-(M_a+M_b)^2}}
{\sqrt{s-(M_a-M_b)^2}+\sqrt{s-(M_a+M_b)^2}}
    +i\bigg]\,,\nn \\
    \beta&=&-\frac{M_a^2-M_b^2}{16\pi^2}
    \ln\frac{M_a}{M_b}.
    \eea
The point $s=(M_a+M_b)^2$ is singular.
For $s<(M_a+M_b)^2$ we write
$\sqrt{s-(M_a+M_b)^2}\to i\sqrt{(M_a+M_b)^2-s}$,
the points $s=(M_a-M_b)^2$ and $s=0$
are not singular, the pole singularity at $s=0$
is cancelled due to the term with $\beta$.

The subtraction constant $b_0$ regulates a value of the meson component near
the threshold, i.e. the fraction of the deuteron-like system.
The zero value of the deuteron-like fraction is realized with
$B_{}(s)\Big|_{s=(M_a+M_b)^2}=0$, namely at: $b_0=0$.

We have eliminated the pole singularity in the loop diagram introducing
the cancellation term $\frac{\beta}{s}$.
Let us remark, however, that the pole singularity in the loop diagram
does not violate the analytical structure of the total amplitude because
the poles in the loop diagram do not lead to new singularities but to zeros
of the amplitude (see Appendix A).

\subsection{One-pole propagator with non-zero orbital momenta
of mesons, $L_{} \neq 0$  }

The loop diagram expression of Eq. (\ref{11})
gives a possibility to write down
analogous terms with non-zero orbital momenta, $L_{\ell} \neq 0$, and
taking into account form factor $G_{\ell}(s)$:
\be
\label{14}
B^{L_{\ell}}_{{\ell}}(s)= k_{\ell}^{2L_{\ell}}G_{\ell}(s)
B_{\ell}(s)G_{\ell}(s)\,.
\ee
Factor $G_{\ell}(s)$ is to be chosen in a form without the
violation of the analytical structure of the amplitude, for example, one
can use the simple exponential form
$G_{\ell}(s)=G_0 \exp(-R^2_\ell s)$.
The exponential form guarantees the convergence of the loop diagrams.
One can use the inverse polynomial function as well:
$ G_{\ell}(s)\sim 1/P_n(s)$ with $P_n(s)=\sum\limits_{\nu=0}^{n} a_\nu  s^\nu $
because zeros of the $P_n(s)$ are not singilar points of the amplitude.

\subsection{Two-pole amplitude}

The two-pole $D$-matrix functions can be written as solutions of the following
equations:
\bea
&&D_{11}\ =\ d_1+D_{11}\,B_{11}\,d_1+D_{12}\,B_{21}\,d_1\,,
\nonumber \\
&&D_{12}\ =D_{11}\,B_{12}\,d_2+D_{12}\,B_{22}\,d_2\,, \nonumber \\
&&D_{21}\ =D_{22}\,B_{21}\,d_1+D_{21}\,B_{11}\,d_1\,, \nonumber \\
&&D_{22}\ =\ d_2+D_{22}\,B_{22}\,d_2+D_{21}\,B_{12}\,d_2\,,
\eea
that results in the explicit form
\bea
&& D_{12} = \frac{d_1 B_{12} d_2}{(1-B_{22}d_2)(1-B_{11}d_1)-d_1 B_{12}\,d_2
B_{21}}\,,
\\
&& D_{11}=\frac{d_1 (1- B_{22}d_2)}{(1-B_{22}d_2)(1-B_{11}d_1)-d_1 B_{12}\,d_2
B_{21}}\,.
\nonumber
\eea
Here $B_{if}=B_{fi}$, zeros of the denominator determine positions of the
poles.
Expressions for $D_{12}$ and $D_{22}$ are given by the replacement of indices
$1\rightleftharpoons 2$. We have $D_{12}=D_{21}$ and a common denominator
for all $D$-functions.

If $B_{12}$ is small we have two separate poles in the region of studies
similar to that discussed in the one-pole case. Non-zero $B_{12}$ means a
mixture of the input pole states and the change of their masses and widths.
At large $B_{if}$ additional poles can appear, the additional poles mean
the appearance of new two-meson states created by mesons of the loop diagrams.

\subsection{D-matrix with an arbitrary number of poles}

The equation for the $D$-matrix can be written as
follows:
\begin{equation}
\hat D(s) =\hat d(s)+ \hat D(s)\hat B(s)\hat d(s) ,
\nonumber
\end{equation}
that gives:
\be  \label{17}
\hat D(s) = \hat d(s)\frac{1}{I-\hat B(s)\hat d(s)}
\ee
where
\bea
\hat D(s)\ &=&\left|
\begin{array}{ccccc}
D_{11}(s) & D_{12}(s) & D_{13}(s) & \cdot & \cdot\\
D_{21}(s) & D_{22}(s) & D_{23}(s) & \cdot & \cdot\\
D_{31}(s) & D_{32}(s) & D_{33}(s) & \cdot & \cdot\\
\cdot     & \cdot     & \cdot     & \cdot & \cdot\\
\cdot     & \cdot     & \cdot     & \cdot & \cdot
\end{array}\right|, \quad
\hat d(s) = \left|
\begin{array}{ccccc}
d_1(s)&0&0&\cdot&\cdot\\
0&d_2(s)&0&\cdot&\cdot\\
0&0&d_3(s)&\cdot&\cdot\\
\cdot&\cdot&\cdot&\cdot&\cdot\\
\cdot     & \cdot     & \cdot     & \cdot & \cdot
\end{array}
\right|,
\nonumber \\
\hat B(s)\ &=&\left|
\begin{array}{ccccc}
B_{11}(s) & B_{12}(s) &B_{13}(s) & \cdot& \cdot\\
B_{21}(s) & B_{22}(s) &B_{23}(s) & \cdot& \cdot\\
     B_{31}(s) &  B_{32}(s)    & B_{33}(s) &\cdot& \cdot\\
     \cdot & \cdot     & \cdot    & \cdot& \cdot\\
     \cdot &  \cdot          & \cdot
\end{array}\right|.
\label{18}
\eea
and
\bea  \label{19}
d_i(s)&=&\frac{1}{m^2_i-s},\\
     B_{if}(s)&=&
\sum\limits_{\ell}
    k^{2L_{\ell}}_{\ell}G_{i\ell}(s)
B_{\ell}(s)G_{\ell f}(s)\, .  \nonumber
\eea
Recall that for the one-pole case
$B_{\ell}(s)\equiv B_{\{ab\}}(s)\equiv B_{}(s)$ is given by
Eq. (\ref{11}).

\section{$D$-function for mesons with spin}

We consider here meson resonances with spin. First, as elucidation examples,
cases with scalar (S), pseudoscalar (P), vector (V), and tensor (T)
particles are considered, after that
the $D$-functions for particles with higher spins are presented.

\subsection{Transition $1^-\to [1^-(k_a) +0^+(k_b)]_{S-wave} \to 1^-$}
To calculate the propagator we should calculate the imaginary part of
the loop diagram and restore the real part using Eq. (\ref{11}).
In this procedure the $S$-wave terms for the transitions
$V^{(in)}\to [V^{(a)}+S^{(b)}]_{S-wave}\to V^{(fin)}$ (see Fig. \ref{f1})
are written as follows:
\bea \label{18}
&&\frac{g^{\perp P}_{\alpha\beta}}{m^2-s}
+\frac{g^{\perp
P}_{\alpha\alpha'}}{m^2-s}G_{(ab)}(s)\int\frac{d^4k_1d^4k_2}{i(2\pi)^4}
\nn \\
&\times&
\frac{\delta(P-k_a-k_b)g^{\perp k_a}_{\alpha'\beta'}}
{(M_a^2-k^2_a-i0)(M_b^2-k^b_2-i0)}\, G_{(ab)}(s)\; \cdot
\frac{g^{\perp P}_{\beta' \beta}}{{m^2-s}}+...
\nn
\\
&=&
\frac{g^{\perp P}_{\alpha\beta}}{m^2-s}
\Big[1+\frac{G_{(ab)}(s)S_V^{VS}(s)B(s)G_{(ab)}(s)}{m^2-s}+...\Big]
\nonumber
\\
&=&
\frac{g^{\perp P}_{\alpha\beta}}{m^2-s-G_{(ab)}(s)S_V^{VS}(s)B(s)G_{(ab)}(s)}.
\eea
Here  $g^{\perp
P}_{\alpha\beta}=g_{\alpha\beta}-\frac{P_{\alpha}P_{\beta}}{P^2}$ and
\bea
3\,S_V^{VS}(s)\ =\ g^{\perp P}_{\alpha\alpha'}   g^{\perp k_a}_{\alpha'\beta'}
g^{\perp P}_{\beta' \alpha}=[2+\frac{(k_aP)^2}{s\,M_a^2}]\,,
\eea
with vectors $k_1$, $k_2$ being the mass-on-shell values ($k_a^2=M^2_a$
, $k_b^2=M^2_b$)
that result:
\be
2(k_aP)=s+M_a^2-M_b^2.
\ee
The width is determined by the imaginary part of the loop diagram, and
    $S_V^{VS}(s)$ is a meromorphic function.

\subsection{Transition $1^-\to [1^-(k_a) +0^-(k_b)]_{P-wave} \to 1^-$ }

For the propagator $V^{(in)}\to [V^{(a)}+\pi^{(b)}]_{P-wave}\to V^{(fin)}$ we
write:
\bea \label{21}
    \frac{g^{\perp P}_{\alpha\beta}}{m^2-s}
&+&\frac{g^{\perp
P}_{\alpha\alpha'}}{m^2-s}G_{(ab)}(s)\int\frac{d^4k_a}{i(2\pi)^4}
\nn \\
&\times&
\,\frac{\epsilon_{\alpha'\gamma'k_aP}\,g^{\perp k_a}_{\gamma'\delta'}\,
\epsilon_{\delta'\beta'Pk_a}}
{(M_a^2-k^2_a-i0)(M_b^2-k_b^2-i0)}\, G_{(ab)}(s)\,
\frac{g^{\perp P}_{\beta' \beta}}{{m^2-s}}+...
\nonumber
\\
&=&\frac{g^{\perp P}_{\alpha\beta}}{m^2-s}
\Big[1+\frac{G_{(ab)}(s)S_V^{V\pi}(s)B(s)G_{(ab)}(s)}{m^2-s}+...\Big]
\nonumber
\\
&=&\frac{g^{\perp
P}_{\alpha\beta}}{m^2-s-G_{(ab)}(s)S_V^{V\pi}(s)B(s)G_{(ab)}(s)},
\eea
with spin factor $S_V^{V\pi}(s)$ determined by
    mass-on-shell mesons, $k_a^2=M^2_a$, $\,k_b^2=M^2_b$:
\bea
3\,S_V^{V\pi}(s)=g^{\perp P}_{\alpha\alpha'} \epsilon_{\alpha'\gamma'k_aP}\,
g^{\perp k_a}_{\gamma'\delta'}\,
\epsilon_{\delta'\beta'Pk_a}
g^{\perp P}_{\beta' \alpha}=
\frac{2s}{M_aM_b}\Big[\frac{(k_aP)^2}{s}-M_a^2\Big]\,.
\eea
Recall that the loop diagram factor $B(s)$ is given in Eq. (\ref{11}).

\subsection{ The $S$-wave transition of vector-axial state \\
$1^+\to [1^-(k_a)+1^-(k_b)]_{S-wave} \to 1^+$ }

For the transition $A^{(in)}\to [V^{(a)}+V^{(b)}]_{S-wave} \to A^{(fin)}$
the second diagram of the set of Fig. \ref{f1} reads:
\bea \label{}
&&
\frac{g^{\perp P}_{\alpha\alpha'}}{m^2-s}
i\epsilon_{\alpha'\gamma'\gamma'' P}\;  \cdot G_{(ab)}(s)
\int\frac{d^4k_ad^4k_b}{i(2\pi)^4}\delta(P-k_a-k_b)\,
\\
&&
\times\frac{g^{\perp k_a}_{\gamma'\delta'} g^{\perp k_b}_{\gamma''\delta''}}
{(M_a^2-k^2_a-i0)(M_b^2-k^2_b-i0)}\, G_{(ab)}(s)\; \cdot
(-i)\epsilon_{\delta'\delta''\beta' P}
\frac{g^{\perp P}_{\beta' \beta}}{{m^2-s}}
\nonumber
\eea
with the notation
$\epsilon_{\delta'\delta''\beta' P}=\epsilon_{\delta'\delta''\beta'\beta''}
P_{\beta''}$
and $g^{\perp P}_{\alpha\beta} =g_{\alpha\beta}-\frac{P_\alpha P_\beta}
{P^2}$.

The spin factor of the second term is equal to:
\bea  \label{20}
g^{\perp P}_{\alpha\alpha'}
\epsilon_{\alpha'\gamma'\gamma'' P}\;  \cdot
g^{\perp k_a}_{\gamma'\delta'}g^{\perp k_b}_{\gamma''\delta''}
    \cdot
\epsilon_{\delta'\delta''\beta' P}\cdot
g^{\perp P}_{\beta' \beta}\ =\ g^{\perp P}_{\alpha\beta}\,
S_{A}^{V_aV_b}(s)\,.
\eea

Let us remind that here we mean
$M_a^2=k^2_a$ an $M_b^2=k^2_b$.
The resonance propagator is written as follows:
\be  J=1:\qquad
\frac{g^{\perp
P}_{\alpha\beta}}{m^2-s-G_{(ab)}(s)S_{A}^{V^{(a)}V^{(b)}}(s)B(s)G_{(ab)}(s)}\,.
\ee

\subsection{ The $S$-wave transitions  $S\to V^{(a)}+V^{(b)} \to S$, \\
    and $T\to V^{(a)}+V^{(b)} \to T$}

The propagator for resonance with $J=0$ is:
\bea
&&  J=0:\qquad
\frac{1}{m^2-s-G_{(ab)}(s)S_{S}^{V^{(a)}V^{(b)}}(s)B(s)G_{(ab)}(s)}\,,
\\
&&
    S_{S}^{V^{(a)}V^{(b)}}(s) =
\frac13
G^2(s)\Gamma_{0}^{\gamma'\gamma''}(\perp P)\;  \cdot
O_{\gamma'}^{\delta'}({\perp k_a})O^{\gamma''}_{\delta''}({\perp k_b})
    \cdot
\Gamma^{\delta'\delta''}_{0}({\perp P})
\,,
\nonumber  \\
&&
\Gamma_{0}^{\delta'\delta''}(\perp P)=
O_{\delta'}^{\delta''}({\perp P})
\,.
\nonumber
\eea
Here we introduce the vertex function $\Gamma_{0}^{\gamma'\gamma''}(\perp P)$
and denote
$O_{\gamma''}^{\gamma'}({\perp P})=g_{\gamma''\gamma'}^{\perp P}$,
see Appendix B for details.

We write for $J=2$:
\bea
&&  J=2:\qquad
\frac{O^{\alpha_1\alpha_2}_{\beta_1\beta_2}({\perp
P})}{m^2-s-G^{(ab)}(s)S_{T}^{V^{(a)}V^{(b)}}(s)B(s)G^{(ab)}(s)}\,,
\\
&&
S_2(s,k_1^2,k_2^2) =
\frac15 G^2(s)
O_{\alpha_1\alpha_2}^{\gamma'\gamma''}(\perp P)\;  \cdot
O_{\gamma'}^{\delta'}({\perp k_1})O_{\gamma''}^{\delta''}({\perp k_2})
    \cdot
O_{\delta'\delta''}^{\alpha_1\alpha_2}({\perp P})\,.
\nonumber
\eea
The operator for the tensor state,
$O_{\alpha_1\alpha_2}^{\gamma'\gamma''}(\perp P)$,
is given in  Appendix B.

\section{Baryon resonances }
Propagators for baryon resonances can be constructed in a way
analogous to that for mesons but with some complication, namely:
determining one-channel rescattering we face two
basic loop functions, $B(s)$ and $ \widetilde B(s)$.

First, we present several examples of propagators for spin-1/2 resonances.
Then the cases with larger spins ($J>1/2$) are discussed. The technique used
here for fermions with spins $J>1/2$ is given in Appendix C.

\subsection {Spin-$1/2$ state and its decay with the emission of a scalar
meson}

We consider here transitions of the type $N^*(\frac12^+)\to
[S(0^+)+N(\frac12^+)]\to
N^*(\frac12^+)$.

The first two terms of the series shown in Fig. \ref{f1} are written as:
\bea  &&
\qquad
    \frac{\hat P+\sqrt{s}}{m^2-s}+\frac{\hat P+\sqrt{s}}{m^2-s}
\cdot G(s)\,B^{}_{N^*}(s)\,G(s)\cdot\frac{\hat P+\sqrt{s}}{m^2-s}\,,
\eea
where the loop function $B^{SN}_{N^*}(s)=B^{}_{N^*}(s)\cdot 2\sqrt s$ has the following
form:

\bea
B^{SN}_{N^*}(s)
&&=\int\frac{d^4k}{i(2\pi)^4}
\frac{\hat k+M_N}{(k^2-M_N^2-i0)((P-k)^2-M_S^2-i0)}
\cdot(\hat P+\sqrt{s})
\nonumber\\
=\int\frac{d^4k}{i(2\pi)^4}&&
\frac{\frac{(kP)}{P^2}\hat P+M_N}{(k^2-M_N^2-i0)((P-k)^2-M_S^2-i0)}
\cdot(\hat P+\sqrt{s})
\nonumber
\\
=\int\frac{d^4k}{i(2\pi)^4}&&
\frac{\frac{(kP)}{s}\sqrt s+M_N}{(k^2-M_N^2-i0)((P-k)^2-M_S^2-i0)}
\cdot 2\sqrt{s} \,.
\eea
We use $k=k^\perp+\frac{(kP)}{P^2} P$ and
$(\hat P+\sqrt{s})(A\hat P+B)(\hat P+\sqrt{s})=
(\hat P+\sqrt{s})(A\sqrt{s}+B)\cdot 2\sqrt{s}$;
recall
that the loop diagram hadrons are mass-on-shell in the imaginary part, and
$2(kP)=s+M_N^2-M_S^2$.

The propagator for the $N^*(\frac12^+)$-state reads:
\bea
\frac{\hat P+\sqrt{s}}{m^2-s-G^2(s)\,B^{S N}_{N^*}(s)}\,,
\eea
with loop function $B^{S N}_{N^*}(s)$:
\bea
B^{S N}_{N^*}(s)=2[(kP)B(s)+M_N^2 \widetilde B(s)]\,.
\eea
The loop function $B(s)$ given in Eq. (\ref{11}).
The new basic term $\tilde B(s)$ reads as follows:
\bea \label{30}
\widetilde B(s)&=& \tilde b_0 +
\frac{\sqrt{[s-(M_S+M_N)^2][s-(M_S-M_N)^2]}}{16\pi M_N\sqrt{ s}}
\\
&&
\times\bigg[\frac{1}{\pi}\ln
\frac{\sqrt{s[s-(M_S-M_N)^2]}-\sqrt{(M_N-M_S)^2[s-(M_S+M_N)^2]}}
{\sqrt{s[s-(M_S-M_N)^2]}+\sqrt{(M_N-M_S)^2[s-(M_S+M_N)^2]}}
    +i\bigg]\,.
\nonumber
    \eea
Singularities  $s=0$ and $s=(M_S-M_N)^2$ are absent in $\widetilde B(s)$,
the only present singularity is the threshold one $s=(M_S+M_N)^2$.
In the determination of the $\widetilde B(s)$ an uncertainty exists which is
related to zeros of the loop functions; this item is discussed in Appendix A,
subsection 7.3.

At $(M_S-M_N)^2=0$ the loop function has a simple form:
\be \label{33}
\widetilde B(s)= \tilde b_0 +i
\frac{\sqrt{[s-(M_S+M_N)^2]}}{16\pi M_N}\,.
\ee

\subsection {Decay of the $N^*(\frac12^+)$-state with the emission of a
pseudoscalar meson}

The propagator for the $N^*(\frac12^+)$-state with
the transition $N^*(\frac12^+)\to \pi(0^-)+N(\frac12^+)\to
   N^*(\frac12^+)$
taken into account can be written as:
\be
\frac{\hat P+\sqrt{s}}{m^2-s-G^2(s)B^{\pi  N}_{N^*}(s)}\,, \\
\ee
where $B^{\pi  N}_{N^*}(s)$ is determined by the loop diagram
$N^*(\frac12^+)\to \pi(0^-)+N(\frac12^+)\to N^*(\frac12^+)$, namely:
\bea  &&
B^{\pi N}_{N^*}(s)=
\\ \nonumber
&&=\int\frac{d^4k}{i(2\pi)^4}
\frac{ i\hat k_\perp\gamma_5(\hat k+M_N) i\gamma_5\hat k_\perp}{(k^2-M_N^2-i0)
  ((P-k)^2-M_\pi^2-i0)}
\cdot(\hat P+\sqrt{s})
\\ \nonumber
&&
=\int\frac{d^4k}{i(2\pi)^4}
\frac{-k^2_\perp(\frac{(kP)}{P^2}\hat P+M_N)}{(k^2-M_N^2-i0)((P-k)^2-M_\pi
^2-i0)} \cdot(\hat P+\sqrt{s})
\\ \nonumber
&&
=\int\frac{d^4k}{i(2\pi)^4}
\frac{(\frac{(kP)^2}{P^2} -M_N^2)(\frac{(kP)}{P^2}\sqrt
s+M_N)}{(k^2-M_N^2-i0)((P-k)^2-M_\pi ^2-i0)} \cdot 2\sqrt{s} \,. \eea Recall, we use
$k=k^\perp+\frac{(kP)}{P^2} P$ and $\hat P\to \sqrt s$.

The hadron rescattering factor is
\be
B^{\pi  N}_{N^*}(s)=2(\frac{(kP)^2}{P^2} -M_N^2)\Big[(kP)B(s)+M_N^2
\widetilde B(s)\Big]\,,
\ee
with $B(s)$ and $\widetilde B(s)$ given in Eqs. (\ref{11}) and (\ref{30}).

\subsection {Transitions $\Delta(\frac32^+) \to [N(\frac12^+)+\pi(0^-)]
\to \Delta(\frac32^+) $}

The propagator of the $\Delta(\frac32^+)$-resonance,
taking into account the transition
    $\Delta(\frac32^+) \to N(\frac12^+)\pi(0^-)$
(see Chapter 5 of ref. [\cite{book3}] and references therein) reads:
\be
    \frac{(-g_{\mu\nu}^\perp +\frac13\gamma_{\mu}^\perp \gamma_{\nu}^\perp )
    (\hat P+\sqrt{s})}{m^2-s-G^2(s)B^{\pi N}_{\Delta}(s)}\,, \\
\ee
see Appendix C for details.
The factor $B^{\pi  N}_{\Delta}(s)$ is determined by the $P$-wave loop diagram
$\Delta(\frac32^+) \to [N(\frac12^+)\pi(0^-)]_{P-wave}\to \Delta(\frac32^+)$,
namely:
\bea
&&
(-g_{\mu\nu}^\perp +\frac13\gamma_{\mu}^\perp \gamma_{\nu}^\perp )\,
 (\hat P+\sqrt{s})\,B^{\pi N}_{\Delta}(s)
\\
\nonumber
&&
=(-g_{\mu\mu'}^\perp +\frac13\gamma_{\mu}^\perp \gamma_{\mu'}^\perp
)(\hat P+\sqrt{s})\int\frac{d^4k}{i(2\pi)^4}
   \frac{  k^\perp_{\mu'}(\hat k+M_N)
 k^\perp_{\mu'}}{(k^2-M_N^2-i0)((P-k)^2-M_\pi
^2-i0)}
\\ \nonumber
&&\times
(-g_{\nu'\nu}^\perp +\frac13\gamma_{\nu'}^\perp \gamma_{\nu}^\perp )\,(\hat
P+\sqrt{s})
\\ \nonumber
&&= (g_{\mu\nu}^\perp -\frac13\gamma_{\mu}^\perp
\gamma_{\nu}^\perp )(\hat P+\sqrt{s}) \int\frac{d^4k}{i(2\pi)^4}
\frac{k_\perp^2\Big(\frac{(kP)}{P^2}\hat P+M_N\Big)}{(k^2-M_N^2-i0)
((P-k)^2-M_\pi^2-i0)}2\sqrt{s}
\,.
\eea
Therefore the $\pi N$ rescattering factor
can be written as ($k_\perp^2=M_N^2-\frac{(Pk)^2}{P^2}$):
\be
B^{\pi  N}_{\Delta}(s)=2\Big(-M_N^2+\frac{(Pk)^2}{s}\Big)\Big[(kP)B(s)+M_N^2
\widetilde B(s)\Big]\,,
\ee
with $B(s)$ and $\widetilde B(s)$ given in eqs. (\ref{11}) and (\ref{30}).

\section{Deuteron-like component}
Let us consider in a more detailed way the case when the pole singularity
is located near the threshold, $m\simeq M_a+M_b$.
In this situation the deuteron-like component in the resonant state
manifests itself evidently.

As an example we consider a case of the one-channel and one-pole
amplitude, leading to $S$-wave decay-products.
The scattering amplitude $ab\to ab$ reads:
\be
\frac{1}{k}e^{i\delta}\sin\delta=
\frac{G^2}{m^2-s-G^2(B_\Re(s)+ik)}
\ee
where $B_\Re(s)$ is the real part of the loop diagram.
Expanding this amplitude in a series over relative
momentum of mesons, $k$, one has:
\bea \label{38}
&&
\frac{G^2}{[m^2-(M_a+M_b)^2-G^2B_t]-k^2\frac{(M_a+M_b)^2}{M_aM_b}(1+G^2
B'_t)-ikG^2}
\\
&&
= \frac{a_0}{1+\frac12a_0r_0k^2 -ia_0k} \,,
\nonumber
\eea
here   $B_t=B_\Re (s)_{s=(M_a+M_b)^2}$ and
$ B'_t=\Big(\frac{d B_\Re (s)}{ds} \Big)_{s=(M_a+M_b)^2}$,
whereas $a_0$ is the scattering length and $r_0$ is the
effective radius of the $ab$-system:
\bea && a_0=\frac{G^2}{m^2-(M_a+M_b)^2-G^2B_t}\,,\\
&&
r_0=-2\frac{(M_a+M_b)^2}{M_aM_b}(G^{-2}+ B'_t) \,.
\nonumber
\eea
At large negative $a_0$ the system has a stable component (an analog of the
deuteron), at positive $a_0$ the resonance signal appears only in the
continuous spectrum (the system is the analog of the singlet state in $pp$).
A small value of $[m^2-(M_a+M_b)^2-G^2B_t]$ (a large value of $|a_0|$) can
exist independently of details of the long-range hadron-hadron
interaction.

The large density of the levels in multi-particle systems
enlarges the probability to face the effect of appearance of the
deuteron-like components.

\section{Conclusion}

The hadron resonance topic is a key subject for both experimental
studies and theoretical understanding in physics of elementary particles.
An important point in this subject is the correct description of resonances.
Using the language of hadron amplitudes this means a correct representation
of the analytical structure of amplitudes. First of all, it concerns the
propagators of resonances.

The experimental study of resonances is connected mainly to the investigation
of multi-hadron reactions, the simplest reactions are three-particle ones.
The description of Dalitz-plot data faces problems with the simultaneous
presentation of resonances from different channels and the incorporation
of requirements of the unitarity and analyticity into phenomenological
analyses.

The use of three-body equations leads to implementing the analyticity and
the three-body unitarity into the amplitude; in the non-relativistic case
such an implementation can be performed using the Faddeev equation
\cite{faddeev} while for the relativistic consideration the dispersion
relation technique looks as the most appropriate one. In this case the
resonance propagators with included decay components are essential. But in
the first attempts to write dispersion relation equations \cite{tre60}
problems appeared in choosing the way of integration.

A correct integration over a three-body intermediate state was performed
in Ref.~[\citen{aag62}], the corresponding consequences of such an
integration are discussed already for a long
time \cite{ad66,ait66,ait2012,book4}.
A realistic system of equations for coupled channel amplitudes for
proton-antiproton annihilation at rest
$[p\bar p]_{at\, rest}\to \pi\pi\pi,\;\eta\eta\pi,\;K\bar K\pi$
was written in Ref.~[\citen{ava-3body}]
(see also Ref.~[\citen{book4}], Chapter 5). A critical issue in the
equations is the dispersion relation presentation of two-meson amplitudes
and the corresponding resonances.

The multi-component structure of the constructed propagators for
resonances allows to fix deuteron-like states.  Examples are presented by
states with hidden charm. There are several candidates for states with
long-distant hadronic components:
$X(3872)\to J/\Psi \pi\pi$ (nearby threshold $\bar DD^*$) [\citen{X3872}] ,
$X(3900)\to J/\Psi \pi$ (nearby threshold $\bar DD^*$) [\citen{X3900}],
$X(4020)\to J/\Psi \pi$ (nearby threshold $\bar D^*D^*$) [\citen{X4020}].
A popular interpretation of these states is that they are meson-meson molecules
( $\bar DD^*$ and  $\bar D^*D^*$). But it is possible that the states have two
components, namely, short-range and long-range ones. That happens when a
quark-gluon state (presumably a short-range one) is situated (may be
accidentally) in the vicinity of the decay threshold.

To conclude: the construction of propagators of composite states with
decay loop diagrams taken into account is a relevant subject for both
experimental and theoretical studies in
hadron physics.

\subsection*{Acknowledgement}

We thank D.I. Melikhov for useful comments.
The paper was supported by grant RSF 16-12-10267.

\section*{Appendix A: Loop diagram analyticity}

The convergence of the loop diagram, $B(s)$, can be guaranteed by
introducing the vertex $s$-dependence or using the subtraction procedure:
\be  \label{42}
\int\limits_{(M_a+M_b)^2}^{+\infty}\frac{ds'}{\pi}
\frac{\rho_{\alpha'}(s')}{s'-s-i0} \to  B(s=s_0) +
\int\limits_{(M_a+M_b)^2}^{+\infty}\frac{ds'}{\pi}
\frac{\rho_{\alpha'}(s')}{s'-s-i0}\cdot \frac{s-s_0}{s'-s_0}\,.
\ee
Here the subtraction procedure is used.
In our studies we put $s_0=(M_a+M_b)^2$.

We face two types of the imaginary parts for the loop diagrams:
    \bea
&&
Im\; B(s)=
\frac{\sqrt{[s\!-\!(M_a\!+\!M_b)^2][s\!-\!(M_a\!-\!M_b)^2]}}{16\pi s}\,,
\\
&&
Im\; \widetilde B_{}(s)=
\frac{\sqrt{[s\!-\!(M_S\!+\!M_N)^2][s\!-\!(M_S\!-\!M_N)^2]}}{16\pi M_N \sqrt
s}\,.
\nonumber
\eea
   Within these
imaginary parts we restore the loop diagrams, see Eqs. (\ref{11}) and
(\ref{30}).

\subsection*{Meson-meson loop diagram}

Let us consider the analytical sructure of the $ B(s)$ in a more detailed
way, keeping $s_0=(M_a+M_b)^2$.

Below the threshold, at $(M_a-M_b)^2<s<(M_a+M_b)^2$, the loop diagram reads:
\bea
\label{12}
B_{}(s)&=&b_0\!+\! \beta\frac{(M_a+M_b)^2-s}{s(M_a+M_b)^2}
\!+\!i\frac{\sqrt{[-s\!+\!(M_a\!+\!M_b)^2][s\!-\!(M_a\!-\!M_b)^2]}}{16\pi
   s}
\nn \\ &\times&
\bigg[\frac{1}{\pi}\ln
\frac{\sqrt{s-(M_a\!-\!M_b)^2}-i\sqrt{-s+(M_a\!+\!M_b)^2}}
{\sqrt{s\!-\!(M_a\!-\!M_b)^2}+i\sqrt{-s\!+\!(M_a\!+\!M_b)^2}}
    \!+\!i\bigg]
\nn \\
&=&b_0+
\beta\frac{(M_a+M_b)^2-s}{s(M_a+M_b)^2}
+i\frac{\sqrt{[-s\!+\!(M_a\!+\!M_b)^2][s\!-\!(M_a\!-\!M_b)^2]}}{16\pi s}
\nn \\ &\times&
\bigg[-\frac{2i}{\pi}\,\tan^{-1}\bigg(
\frac{\sqrt{-s\!+\!(M_a\!+\!M_b)^2}}{\sqrt{s\!-\!(M_a\!-\!M_b)^2}} \bigg)
+i \bigg] .
    \eea
The last line demonstrates the absence of a singularity in $s=(M_a-M_b)^2$.
Indeed,
in the top-down approach to this point we have:
\bea
&&-\frac{2i}{\pi}\,\tan^{-1}\bigg(
\frac{\sqrt{-s+(M_a+M_b)^2}}{\sqrt{s-(M_a-M_b)^2}} \bigg)+i
\nn \\ &=&-\frac{2i}{\pi}\bigg(
\frac{\pi}{2}- \tan^{-1}
\frac{\sqrt{s-(M_a-M_b)^2}}{\sqrt{-s+(M_a+M_b)^2}} \bigg)+i
\nn \\
&\simeq &
-\frac{2i}{\pi}\bigg(
\frac{\pi}{2}-
\frac{\sqrt{s-(M_a-M_b)^2}}{\sqrt{-s+(M_a+M_b)^2}} \bigg)+i
\eea
with the corresponding cancellation of the singular terms in Eq. (\ref{12}).

\subsection*{Meson-baryon loop diagram}

The meson-nucleon loop diagram $\widetilde B(s)$ below the threshold, at
$(M_S-M_N)^2<s<(M_S+M_N)^2$, reads:
\bea \label{12}
\widetilde B(s)&=&\widetilde b_0\!+\!
i\frac{\sqrt{[-s\!+\!(M_S\!+\!M_N)^2][s\!-\!(M_S\!-\!M_N)^2]}}{16\pi
M_N\sqrt{s}}
\nonumber
\\
\nonumber
&\times&
\bigg[\frac{1}{\pi}\ln
\frac{\sqrt{s[s-(M_S\!-\!M_N)^2]}-i|M_N-M_S|\sqrt{-s+(M_S\!+\!M_N)^2}}
{\sqrt{s[s\!-\!(M_S\!-\!M_N)^2]}+i|M_N-M_S|\sqrt{-s\!+\!(M_S\!+\!M_N)^2}}
    \!+\!i\bigg]
   \nonumber
\\
&=&\widetilde b_0
+i\frac{\sqrt{[-s\!+\!(M_S\!+\!M_N)^2][s\!-\!(M_S\!-\!M_N)^2]}}{16\pi
M_N\sqrt{s}}
\nonumber\\
   &\times&
\bigg[-\frac{2i}{\pi}\,\tan^{-1}\bigg(
\frac{|M_N-M_S|\sqrt{-s+(M_S+M_N)^2}}{\sqrt{s[s-(M_S-M_N)^2]}} \bigg)
+i \bigg] .
   \eea
Near $s= (M_S-M_N)^2$ we write:
\bea
\tan^{-1}\bigg(
\frac{|M_N-M_S|\sqrt{-s+(M_S+M_N)^2}}{\sqrt{s[s-(M_S-M_N)^2]}} \bigg)
\nn \\
=\frac{\pi}{2}
-\tan^{-1}\frac{\sqrt{s[s-(M_S-M_N)^2]}}{|M_N-M_S|\sqrt{-s+(M_S+M_N)^2}} \bigg)
\eea
and the meson-nucleon loop diagram $\widetilde B(s)$ below the threshold is:
\bea
\widetilde B(s)&=&\widetilde b_0
+\frac{\sqrt{[-s\!+\!(M_S\!+\!M_N)^2][s\!-\!(M_S\!-\!M_N)^2]}}{16\pi
M_N\sqrt{s}}
\nonumber
\\
   &&
\times\bigg[-\frac{2}{\pi}\,
\tan^{-1}\frac{\sqrt{s[s-(M_S-M_N)^2]}}{|M_N-M_S|\sqrt{-s+(M_S+M_N)^2}}  \bigg]
.
\label{48}  \eea
It is seen that points $s=0$ and $s=(M_S-M_N)^2$ are non-singular. Moreover,
at $s=(M_S-M_N)^2$
we have $\widetilde B(s)=\widetilde b_0$, see Eq. (\ref{48}), that corresponds
to zero of the $s$-dependent part of the loop diagram. Ambiguities in the
determimation of the loop diagrams are related to zeros of $B(s)$ and
$\widetilde B(s)$.

\subsection*{Ambiguites in the determination of the resonance amplitude }

The ambiguities of the resonance amplitude are due to CDD-poles [\cite{cdd}] .
The resonance amplitude with CDD-poles taken into account is written as
follows:
\be
\frac{B(s)}{1-B(s)+\sum\limits_{n}\frac{\gamma_n}{s-s_n}}
\ee
A redefinition of the type
\be
B(s) \to \frac{ B(s)}{1+\sum\limits_{n}\frac{\gamma_n}{s-s_n} }
\ee
returns us to the used form of amplitudes. But the redefined $B(s)$ differs in
numbers and the positions of zeros.

\section* {Appendix B:  Angular momentum operators for two-meson systems}

We use angular momentum operators
$X^{(L)}_{\mu_1\ldots\mu_L}(k^\perp)$,
$\,Z^{\alpha}_{\mu_1\ldots\mu_L}(k^\perp)$ and the projection operator
$O^{\mu_1\ldots\mu_L}_{\nu_1\ldots\nu_L}(\perp P)$ (see [\cite{book3,book4,AAMMS}]). Let
us recall their definition.

The operators are constructed from the relative momenta
$k^\perp_\mu$ and tensor $g^\perp_{\mu\nu}$. Both of them are
orthogonal to the total momentum of the system:
\be
k^\perp_\mu=\frac12 g^\perp_{\mu\nu}(k_1-k_2)_\nu =k_{1\nu} g^{\perp
P}_{\nu\mu} =-k_{2\nu} g^{\perp P}_{\nu\mu} ,  \qquad
g^\perp_{\mu\nu}=g_{\mu\nu}-\frac{P_\mu P_\nu}{s}\;.
\ee

The operator for $L=0$ is a scalar (we write $X^{(0)}(k^\perp)=1$),
and the operator for $L=1$ is a vector, $X^{(1)}_\mu=k^\perp_\mu $.
The operators $X^{(L)}_{\mu_1\ldots\mu_L}$ for $L\ge 1$ can be written
in the form of a recurrency relation:
\bea
X^{(L)}_{\mu_1\ldots\mu_L}(k^\perp)&=&k^\perp_\alpha
Z^{\alpha}_{\mu_1\ldots\mu_L}(k^\perp)\equiv  k^\perp_\alpha
Z_{\mu_1\ldots\mu_L,\alpha}(k^\perp)  ,
\nonumber\\
Z^{\alpha}_{\mu_1\ldots\mu_L}(k^\perp)&\equiv &
Z_{\mu_1\ldots\mu_L,\alpha}(k^\perp)=
\frac{2L-1}{L^2}\Big (
\sum^L_{i=1}X^{{(L-1)}}_{\mu_1\ldots\mu_{i-1}\mu_{i+1}\ldots\mu_L}(k^\perp)
g^\perp_{\mu_i\alpha}-
\nonumber \\
    -\frac{2}{2L-1}  \sum^L_{i,j=1 \atop i<j}
&g^\perp_{\mu_i\mu_j}&
X^{{(L-1)}}_{\mu_1\ldots\mu_{i-1}\mu_{i+1}\ldots\mu_{j-1}\mu_{j+1}
\ldots\mu_L\alpha}(k^\perp) \Big ).
\label{Vz}
\eea
    We have a convolution equality
$X^{(L)}_{\mu_1\ldots\mu_{L}}(k^\perp)k^\perp_{\mu_L}=k^2_\perp
X^{(L-1)}_{\mu_1\ldots\mu_{L-1}}(k^\perp)$, with $k^2_\perp\equiv
k^\perp_{\mu}k^\perp_{\mu}$, and the tracelessness property of
$X^{(L)}_{\mu\mu\mu_3\ldots\mu_{L}}=0$. On this basis, one can write
down the normalization condition for orbital angular operators:
\bea
\int\frac{d\Omega}{4\pi}
X^{(L)}_{\mu_1\ldots\mu_{L}}(k^\perp)X^{(L)}_{\mu_1\ldots\mu_{L}}
(k^\perp)
    = \alpha_L k^{2L}_\perp \; ,\quad
\alpha_L\ =\ \prod^L_{l=1}\frac{2l-1}{l}  ,
\label{Valpha}
\eea
    where the integration is performed over spherical variables
$\int d\Omega/(4\pi)=1$.

Iterating Eq. (\ref{Vz}), one obtains the
following expression for the operator $X^{(L)}_{\mu_1\ldots\mu_L}$
at $L\ge 1$:
\bea
\label{Vx-direct}
&&X^{(L)}_{\mu_1\ldots\mu_L}(k^\perp)=
\alpha_L \bigg [
k^\perp_{\mu_1}k^\perp_{\mu_2}k^\perp_{\mu_3}k^\perp_{\mu_4}
\ldots k^\perp_{\mu_L}- \\
&&-\frac{k^2_\perp}{2L-1}\bigg(
g^\perp_{\mu_1\mu_2}k^\perp_{\mu_3}k^\perp_{\mu_4}\ldots
k^\perp_{\mu_L}
+g^\perp_{\mu_1\mu_3}k^\perp_{\mu_2}k^\perp_{\mu_4}\ldots
k^\perp_{\mu_L} + \ldots \bigg)+
\nonumber \\
&&+\frac{k^4_\perp}{(2L\!-\!1)(2L\!-\!3)}\bigg(
g^\perp_{\mu_1\mu_2}g^\perp_{\mu_3\mu_4}k^\perp_{\mu_5}
k^\perp_{\mu_6}\ldots k^\perp_{\mu_L}
\nn \\
&&+
g^\perp_{\mu_1\mu_2}g^\perp_{\mu_3\mu_5}k^\perp_{\mu_4}
k^\perp_{\mu_6}\ldots k^\perp_{\mu_L}+
\ldots\bigg)+\ldots\bigg ]. \nonumber
\eea
For the projection operators, one has:
\bea
&&\hspace{-6mm}
O= 1 ,\qquad
O^\mu_\nu (\perp P)=g_{\mu\nu}^\perp \, ,
\nonumber \\
&&\hspace{-6mm}O^{\mu_1\mu_2}_{\nu_1\nu_2}(\perp P)=
\frac 12 \left (
g_{\mu_1\nu_1}^\perp  g_{\mu_2\nu_2}^\perp \!+\!
g_{\mu_1\nu_2}^\perp  g_{\mu_2\nu_1}^\perp  \!- \!\frac 23
g_{\mu_1\mu_2}^\perp  g_{\nu_1\nu_2}^\perp \right ).
\eea
For higher states, the operator can be calculated using the
recurrent expression:
\bea
&&O^{\mu_1\ldots\mu_L}_{\nu_1\ldots\nu_L}(\perp P)=
\frac{1}{L^2} \bigg (
\sum\limits_{i,j=1}^{L}g^\perp_{\mu_i\nu_j}
O^{\mu_1\ldots\mu_{i-1}\mu_{i+1}\ldots\mu_L}_{\nu_1\ldots
\nu_{j-1}\nu_{j+1}\ldots\nu_L}(\perp P)- \nonumber
    \\
&&- \frac{4}{(2L-1)(2L-3)} \times \sum\limits_{i<j\atop k<m}^{L}
g^\perp_{\mu_i\mu_j}g^\perp_{\nu_k\nu_m}
O^{\mu_1\ldots\mu_{i-1}\mu_{i+1}\ldots\mu_{j-1}\mu_{j+1}\ldots\mu_L}_
{\nu_1\ldots\nu_{k-1}\nu_{k+1}\ldots\nu_{m-1}\nu_{m+1}\ldots\nu_L}(\perp
P) \bigg ).
\eea
The projection operators obey the relations:
\bea
O^{\mu_1\ldots\mu_L}_{\nu_1\ldots\nu_L}(\perp P)
X^{(L)}_{\nu_1\ldots\nu_L}(k^\perp)&=&
X^{(L)}_{\mu_1\ldots\mu_L}(k^\perp)\, ,\nonumber \\
O^{\mu_1\ldots\mu_L}_{\nu_1\ldots\nu_L}(\perp P) k_{\nu_1} k_{\nu_2}
\ldots k_{\nu_L} &=& \frac
{1}{\alpha_L}X^{(L)}_{\mu_1\ldots\mu_L}(k^\perp) .
\eea
    Hence, the product of the two $X^L(k_\perp)$ operators results in the
Legendre polynomials as follows:
\be
X^{(L)}_{\mu_1\ldots\mu_L}(k_1^\perp) (-1)^L
O^{\mu_1\ldots\mu_L}_{\nu_1\ldots\nu_L}(\perp P)
X^{(L)}_{\nu_1\ldots\nu_L}(k_2^\perp)\!=\!\alpha_L
\Big(\sqrt{-k_1^{\perp 2}}\sqrt{-k_2^{\perp 2}}\Big)^L P_L(z),
\ee
where $z\equiv (-{ k}_{1\nu}^\perp { k}_{2\nu}^\perp)/(
\sqrt{-k_1^{\perp 2}}\sqrt{-k_2^{\perp 2}})$.

\section*{Appendix C: Baryon resonances, wave functions and propagators}
We construct spin-dependent propagators which do not change their spin
structure with the inclusion of the loop-diagram interaction. The corresponding
spin wave functions are eigenfunctions for the interaction. In the framework of
this procedure we work with the effective mass of the system, and this
effective mass depends on the energy, $M(s)$.
For resonance systems we write $M^2(s)=s$, for detail see [\cite{book3,book4,A,AS}].

\subsection*{Baryon spin-1/2 wave function}
The spin-dependent numerator of the $D$-function reads:
\begin{eqnarray} \label{VTB12}
&& \sum_{j=1,2}\, \psi_{j}(p)\,\bar \psi_{j}(p)\ =\
\hat p+M(s) , \quad
    \sum_{j=3,4}\, \psi_{j}(p)\, \bar \psi_{j}(p)\ =\
-(\hat p+M(s))_{}\ .
\end{eqnarray}
where $M(s)$ is the effective mass of the resonance system. It means that
we work  with baryon wave functions $\psi (p)$ and
$\bar \psi(p)=\psi^+(p)\gamma_0$ which obey the following equations for
   spin-1/2 fermions:
    \begin{equation} \label{VTB1}
(\hat p-M(s))\psi(p)=0, \qquad  \bar \psi(p)(\hat
p-M(s))=0,
\end{equation}
Wave functions are normalised as follows:
\begin{eqnarray} \label{VTB11}
    j,j'=1,2&:&\quad \left(\bar \psi_j(p)\psi_{j'}(p)\right)= 2M(s)\
\delta_{jj'},
\nonumber\\
j,j'=3,4&:&\quad \left(\bar \psi_j(p)\psi_{j'}(p)\right)=-2M(s)\ \delta_{jj'}.
\end{eqnarray}
The solution of the equation (\ref{VTB1}) gives us four wave functions:
\begin{eqnarray}\label{VTB4}
j=1,2&:& \quad
\psi_j(p) = \sqrt{p_0+M(s)}\left(\begin{array}{c}
\varphi_j \\ \frac{(\mbox{\boldmath$\sigma p$})}{p_0+M(s)}\ \varphi_j
\end{array}\right),
\nonumber \\
&&\quad \bar \psi_j(p)=
\sqrt{p_0+M(s)}\left(\varphi^+_j, -\varphi^+_j \frac{
(\mbox{\boldmath$\sigma p$})}{p_0+M(s)}\right)\ ,
\nonumber \\
j=3,4&:& \quad \psi_j(-p) =i\sqrt{p_0+M(s)} \left(
\begin{array}{c} \frac{(\mbox{\boldmath$\sigma p$})}{p_0+M(s)}\ \chi_j \\
\chi_j \end{array}\right),
\nonumber \\
&&\quad \bar \psi_j(-p) = -i\sqrt{p_0+M(s)}
\left(\chi^+_j  \frac{ (\mbox{\boldmath$\sigma p$})}{p_0+M(s)}
\,,-\chi^+_j\right),
\end{eqnarray}
where $\varphi_j$ and $\chi_j$ are two-component spinors
normalised as
$ \varphi^+_j \varphi_{j'}
=\delta_{jj'}$ and $  \chi^+_j \chi_{j'} =\delta_{jj'}$ .

Solutions with $j=3,4$ refer to antibaryons. The corresponding wave
function is defined as
\be \label{VTB7} j=3,4: \quad \psi^c_j(p)\ =\
C\bar \psi^T_j(-p)\ ,\quad C^{-1}\gamma_\mu C=-\gamma_\mu ^T\ .
\ee
We see that
    $\psi^c_j(p)$ satisfies the equation:
\be \label{VTB8b}
(\hat p-M(s))\psi^c_j(p) =0 \ .
\ee

\subsection*{Spin-$\frac 32$ wave functions}

To describe resonance states $\Delta $ and $\bar\Delta $, we use the wave
functions $\psi_\mu (p)$ and $\bar \psi_\mu(p)=\psi^+_\mu(p)\gamma_0$
which satisfy the following constraints:
    \bea \label{VD1}
    &&(\hat p-M(s))\psi_\mu(p)=0, \qquad \bar \psi_\mu(p)(\hat p-M(s))=
0,\nonumber \\
&&p_\mu \psi_\mu(p)=0, \qquad         \gamma_\mu \psi_\mu(p)=0\ .
    \eea
Here $\psi_\mu(p)$ is a four-component spinor and $\mu$ is a
four-vector index. Sometimes, to underline spin variables, we use
    the notation $\psi_\mu(p;j)$.

\subsubsection*{Wave function for  $\Delta$}

The  equation (\ref{VD1}) gives four wave functions for the
$\Delta$:
\bea \label{VD2}
j=1,2:&& \psi_\mu(p;j) =
\sqrt{p_0+M(s)}\left(\begin{array}{c} \varphi_{\mu\perp}(j) \\
\frac{(\mbox{\boldmath$\sigma p$})}{p_0+M(s)}\ \varphi_{\mu\perp}(j)
\end{array}\right),
    \nonumber\\
&& \bar \psi_{\mu}(p;j)=
\sqrt{p_0+M(s)}\left(\varphi^+_{\mu\perp}(j), -\varphi^+_{\mu\perp}(j) \frac{
(\mbox{\boldmath$\sigma p$})}{p_0+M(s)}\right),
\eea
where the two-component spinors $\varphi_{\mu\perp}(a)$ are determined to be
perpendicular to $p_\mu$
thus keeping for $\Delta $ four independent spin components
$\mu_z=3/2,1/2,-1/2,-3/2$ related to the spin $S=3/2$ and removing
    the components with $S=1/2$.

The completeness conditions for the spin-$\frac 32$ wave functions can
be written as follows:
\begin{eqnarray} \label{VD6}
    \sum_{j=1,2}\,
\psi_{\mu}(p;j)\,\bar \psi_{\nu}(p;j)
&=&
(\hat p+M(s)) \left
(-g^\perp_{\mu\nu}+\frac 13 \gamma^\perp_{\mu} \gamma^\perp_{\nu}
\right )\nonumber\\
&=&
(\hat p+M(s))\frac 23 \left
(-g^\perp_{\mu\nu}+\frac 12 \sigma^\perp_{\mu\nu}
\right )  \  ,
\end{eqnarray}
where $g^\perp_{\mu\nu}\equiv g^{\perp p}_{\mu\nu}$ and
$\gamma^\perp_{\mu}= g^{\perp p}_{\mu\mu'}\gamma_{\mu'}$. The
factor $(\hat p+M(s))$  commutates with
$(g^\perp_{\mu\nu}-\frac 13\gamma^\perp_{\mu}\gamma^\perp_{\nu})$ in
(\ref{VD6}) because
    $\hat p\gamma^\perp_{\mu}\gamma^\perp_{\nu}=
\gamma^\perp_{\mu}\gamma^\perp_{\nu}\hat p$. The matrix
$\sigma^\perp_{\mu\nu} $ is determined in a standard way,
$\sigma^\perp_{\mu\nu}=\frac 12(\gamma^\perp_{\mu}\gamma^\perp_{\nu}-
\gamma^\perp_{\nu}\gamma^\perp_{\mu})$.

\subsubsection*{Wave function for $\bar \Delta$}

The anti-delta, $\bar \Delta$, is determined by the following four wave
functions:
\bea \label{VD7}
j=3,4: && \psi_\mu(-p;j) =i\sqrt{p_0+M(s)} \left(
\begin{array}{c} \frac{(\mbox{\boldmath$\sigma p$})}{p_0+M(s)}\
\chi_{\mu\perp}(j) \\
\chi_{\mu\perp}(j) \end{array}\right),\nonumber \\
&& \bar \psi_\mu(-p;j) =
-i\sqrt{p_0+M(s)} \left(\chi^+_{\mu\perp}(j)  \frac{
(\mbox{\boldmath$\sigma p$})}{p_0+M(s)} \,,-\chi^+_{\mu\perp}(j)\right)\,.
\eea
The completeness conditions for spin-$\frac 32$ wave functions  with
$j=3,4$ are
\begin{eqnarray}
\label{VD9}
    \sum_{j=3,4}\, \psi_{\mu}(-p;j)\, \bar \psi_{\nu}(-p;j)& =&
-(\hat p+M(s)) \left (-g^\perp_{\mu\nu}+\frac 13 \gamma^\perp_{\mu}
\gamma^\perp_{\nu} \right )\nonumber\\
&=&  -
(\hat p+M(s))\frac 23 \left
(-g^\perp_{\mu\nu}+\frac 12 \sigma^\perp_{\mu\nu}
\right ).
\end{eqnarray}

The equation (\ref{VD7}) can be rewritten in the form of (\ref{VD2})
    using the charge conjugation matrix $C$ which was introduced for
spin-$\frac 12$ particles.
We write:
\bea \label{VD10}
j=3,4&:& \psi^c_\mu(p;j)\ =\
C\bar \psi^T_\mu(-p;j).
\eea
The wave functions $\psi^c_\mu(p;j)$ with $j=3,4$ obey the equation:
\be \label{VD11}
    (\hat p-M(s))\,\psi^c_\mu(p;j) =0\ .
\ee

\subsubsection*{Projection operators for resonance states with $J>3/2 $.}

The wave function of a resonance state with spin $J=\ell+1/2$, momentum
   $p$ and effective mass term $M(s)$ is given by a tensor four-spinor
$\psi_{\mu_1\ldots\mu_\ell}$. It satisfies the constraints
\be
\label{5-8} (\hat p-M(s))\psi_{\mu_1\ldots\mu_\ell}=0, \quad
p_{\mu_i}\psi_{\mu_1\ldots\mu_\ell}=0,\quad
\gamma_{\mu_i}\psi_{\mu_1\ldots\mu_\ell}=0,
\ee
and the symmetry properties
\bea \label{5-9}
\psi_{\mu_1\ldots\mu_i\ldots\mu_j\ldots\mu_\ell}&=&
\psi_{\mu_1\ldots\mu_j\ldots\mu_i\ldots\mu_\ell}\;,
\nonumber \\
g_{\mu_i\mu_j}\psi_{\mu_1\ldots\mu_i\ldots\mu_j\ldots\mu_\ell}&=&
g^{\perp p}
_{\mu_i\mu_j}\psi_{\mu_1\ldots\mu_i\ldots\mu_j\ldots\mu_\ell}=0 .
\eea
Conditions (\ref{5-8}), (\ref{5-9}) define the structure of the
denominator of the fermion propagator (the projection operator)
which can be written in the following form:
\be \label{5-10}
F^{\mu_1\ldots\mu_\ell}_{\nu_1\ldots \nu_\ell}(p)=(-1)^\ell
(\hat p+M(s)) \Phi^{\mu_1\ldots\mu_\ell}_{\nu_1\ldots \nu_\ell}(\perp p).
\ee
The operator $\Phi^{\mu_1\ldots\mu_\ell}_{\nu_1\ldots
\nu_\ell}(\perp p)$ describes the tensor structure of the propagator.
It is equal to 1 for a ($J=1/2$)-particle and is proportional to
$g^{\perp p}_{\mu\nu}-\gamma^\perp_\mu\gamma^\perp_\nu/3$ for a
particle with spin $J=3/2$ (remind that $\gamma^\perp_\mu=g^{\perp
p}_{\mu\nu}\gamma_\nu$).

The conditions (\ref{5})-(\ref{9}) are identical for fermion and boson
projection operators and therefore the fermion projection operator
can be written as:
\be  \label{5-11}
\Phi^{\mu_1\ldots\mu_\ell}_{\nu_1\ldots \nu_\ell}(\perp p)=
O^{\mu_1\ldots\mu_\ell}_{\alpha_1\ldots \alpha_\ell} (\perp p)
\phi^{\alpha_1\ldots\alpha_\ell}_{\beta_1\ldots \beta_\ell}(\perp p)
O^{\beta_1\ldots \beta_\ell}_{\nu_1\ldots\nu_\ell} (\perp p)\ .
\ee
    The operator $\phi^{\alpha_1\ldots\alpha_\ell}_{\beta_1\ldots \beta_\ell}
(\perp p)$ can be expressed in a rather simple form since all
symmetry and orthogonality conditions are imposed by $O$-operators.
First, the $\phi$-operator is constructed of metric tensors only, which
act in the space of $\perp p$ and $\gamma^\perp$-matrices. Second, a
construction like $ \gamma^\perp_{\alpha_i}\gamma^\perp_{\alpha_j}=
\frac12 g^\perp_{\alpha_i\alpha_j}+\sigma^\perp_{\alpha_i\alpha_j}$
(remind that here $\sigma^\perp_{\alpha_i\alpha_j}=\frac 12
(\gamma^\perp_{\alpha_i}\gamma^\perp_{\alpha_j}-
\gamma^\perp_{\alpha_j}\gamma^\perp_{\alpha_i}$) gives zero if
multiplied by an $O^{\mu_1\ldots\mu_\ell}_{\alpha_1\ldots
\alpha_\ell}$-operator: the first term is due to the traceless
conditions and the second one to symmetry properties. The only
structures which can then be constructed are
$g^\perp_{\alpha_i\beta_j}$ and $\sigma^\perp_{\alpha_i\beta_j}$.
Moreover, taking into account the symmetry properties of the
$O$-operators,  one can use any pair of indices from sets
$\alpha_1\ldots\alpha_\ell$ and $\beta_1\ldots \beta_\ell$, for example,
$\alpha_i\to \alpha_1$ and $\beta_j\to \beta_1$. Then
\be
\phi^{\alpha_1\ldots\alpha_\ell}_{\beta_1\ldots \beta_\ell}(\perp p)=
\frac{\ell+1}{2\ell\!+\!1} \big( g^\perp_{\alpha_1\beta_1}-
\frac{\ell}{\ell\!+\!1}\sigma^\perp_{\alpha_1\beta_1} \big)
\prod\limits_{i=2}^{\ell}g^\perp_{\alpha_i\beta_i} .
\label{5-t1}
\ee
    Since $\Phi^{\mu_1\ldots\mu_\ell}_{\nu_1\ldots \nu_\ell}(\perp p)$ is
determined by convolutions of $O$-operators, see Eq. (\ref{5-11}),
we can replace in (\ref{5-11})
\be \label{5-13}
\hspace{-7mm}
\phi^{\alpha_1\ldots\alpha_\ell}_{\beta_1\ldots \beta_\ell}(\perp p) \to
\phi^{\alpha_1\ldots\alpha_\ell}_{\beta_1\ldots \beta_\ell}( p) =
\frac{\ell+1}{2\ell\!+\!1} \big( g_{\alpha_1\beta_1}-
\frac{\ell}{\ell\!+\!1}\sigma_{\alpha_1\beta_1} \big)
\prod\limits_{i=2}^{\ell}g_{\alpha_i\beta_i} .
\ee
    The coefficients in (\ref{5-13}) are chosen to satisfy the constraints
(\ref{5-8}) and the convolution condition:
\be  \label{5-14}
\Phi^{\mu_1\ldots\mu_\ell}_{\alpha_1\ldots \alpha_\ell}( p) \Phi^{\alpha_1\ldots
\alpha_\ell}_{\nu_1\ldots \nu_\ell}( p)=
\Phi^{\mu_1\ldots\mu_\ell}_{\nu_1\ldots \nu_\ell}( p) \;.
\ee

\end{document}